# DENSITIES OF STATES, MOMENTS, AND MAXIMALLY BROKEN TIME-REVERSAL SYMMETRY

Roger Haydock and C. M. M. Nex,

Department of Physics and Materials Science Institute,

1274 University of Oregon, OR 97403-1274, USA.

Power moments, modified moments, and optimized moments are powerful tools for solving microscopic models of macroscopic systems; however the expansion of the density of states as a continued fraction does not converge to the macroscopic limit point-wise in energy with increasing numbers of moments. In this work the moment problem is further constrained by minimal lifetimes or maximal breaking of time-reversal symmetry, to yield approximate densities of states with point-wise macroscopic limits. This is applied numerically to models with one and two finite bands with various singularities, as well as to a model with infinite band-width, and the results are compared with the maximum entropy approximation where possible.

# 1. Projected Density of States

For complicated macroscopic systems, even if the equations of motion could be solved exactly, the solutions would contain too much information to be useful. When such systems satisfy linear equations of motion, this information can be summarized in the distribution of energies of states, weighted by their projections onto a single degree of freedom, the projected density of states (PDoS).

In most cases macroscopic linear systems cannot be solved exactly, however densities of states for these systems can still be approximated from moments which are averages of polynomials in energy or frequency over the motion of finite subsystems. The difficulty with this approach is extrapolating from the finite subsystems to the thermodynamic limit of the infinite system. In mathematics this is known as the problem of moments [1] which takes the form of constructing a one-dimensional mass distribution (the PDoS) from powers of position averaged over the distribution. In the moment problem the PDoS is expanded as a continued fraction [2] and the thermodynamic limit corresponds to constructing a tail for the continued fraction as discussed further in Sec. 3.

The purpose of this paper is to present a way of evaluating the thermodynamic limit by imposing the physical condition that states of the macroscopic system have minimal lifetimes consistent with the moments, or in other terms, maximal breaking of time-reversal symmetry(MBTS) in the approximate systems. The remainder of this paper discusses the approximation of densities of states for independent electrons in solids; this is quantum mechanics in the Schrödinger picture for independent particles moving in a time-independent potential. In terms of orthonormal, non-stationary orbitals which may be thought of as atomic-like, the equation of motion for a state is,

$$i\hbar\, \partial\Psi/\partial t = \mathbf{H}\, \Psi, \qquad (1)$$

where $\Psi$ is the column vector of coefficients of each orbital in the state, and $\mathbf{H}$ is the Hamiltonian matrix which is Hermitian because the orbitals are orthonormal. The PDoS for a state $\mathbf{u}_0$ in this system is,

$$n_0(E) = |\text{Im}\{R_0(E)\}|/\pi, \qquad (2)$$

where,

$$R_0(E) = \langle \mathbf{u}_0 | (E\mathbf{I}-\mathbf{H})^{-1} | \mathbf{u}_0 \rangle, \qquad (3)$$

the $\mathbf{u}_0$-$\mathbf{u}_0$, matrix element of the resolvent, Im means the imaginary part, $\mathbf{I}$ is the identity matrix, and E is the energy. For E real, $R_0(E)$ only has a non-zero imaginary part where it is singular, at the energies of stationary states of $\mathbf{H}$ [3]. Physically, $n_0(E)$ is the probability density that $\mathbf{u}_0$ is occupied when the system is in a stationary state with energy E.

While this paper addresses independent electrons in the Schrödinger picture, the ideas presented here apply equally to any Hermitian system of linear equations. One example is quantum mechanics in the Heisenberg picture where the orbitals are replaced by localized operators and $\mathbf{H}$ is replaced by commutation with $\mathbf{H}$, the quantum Liouvillian [4]. Another example is classical mechanics in the harmonic approximation where the coefficients of orbitals are replaced by displacements of atoms from equilibrium positions, and $\mathbf{H}$ is replaced by the dynamical matrix[5]. Finally, these ideas apply generally to classical mechanics in the Liouville formulation where functions of the dynamical variables, as well as the operators which act on those functions, obey linear equations of motion[6].

This paper is divided into seven further Secs. The next of these introduces moments as an approach to electronic structure. Section 3 contains a discussion of the problems with approximating the PDoS in solids using moments, and Sec. 4 presents a method for solving these problems using MBTS. In Sec. 5 the convergence properties of the method are developed and illustrated with numerical examples, and the results are compared with the related method of maximum entropy. An MBTS method for single bands is introduced in Sec. 6 and compared with maximum entropy. In Sec. 7 the previous methods are extended to approximation of the physical and second sheets of the full Greenian rather than just the PDoS, and in Sec. 8, the approximations developed in this paper are placed in the context of more general quadratic approximations, together with comments on approximations for multiple bands. A summary and conclusions are in Sec. 9.

**2. Moments**

What makes a quantum system complicated is the number of inequivalent orbitals which are strongly coupled, in the sense that interactions between orbitals are significant compared to the differences in energies between orbitals. In terms of the Hamiltonian matrix **H**, a pair of orbitals is strongly coupled if their off-diagonal element is significant compared with the difference between their diagonal elements.

Large systems are not necessarily complicated. For example a crystal is infinite, but it can be divided into cells which are equivalent under translation symmetries so that a complete set of inequivalent orbitals can be found in a single finite cell. Defects such as impurities, dislocations, or surfaces, make a system qualitatively more complicated by breaking crystalline symmetry so that equivalent cells are no longer finite. Extended systems with disordered potentials or with interactions are additional examples of complicated systems.

For complicated systems, strong coupling prevents perturbation theory from converging, leaving only the variational method for calculating properties of the system. Even the variational principle must be applied with care in order to control the convergence of calculations, and the simplest way to do this is through moments, matrix-elements of functions, usually polynomials, of the Hamiltonian.

In terms of the Hamiltonian matrix $\mathbf{H}$ and an initial state $\mathbf{u}_0$, usually localized and not stationary, the power moments $\{\mu_n\}$ are the matrix-elements $\{<\mathbf{u}_0, \mathbf{H}^n\mathbf{u}_0>\}$ which are the expectation values of powers of energy over the stationary states in $\mathbf{u}_0$. It can be shown [3] that these power moments are also integrals of powers of E over $n_0(E)$. For modified moments, the power $\mathbf{H}^n$ is replaced by a polynomial of degree n in $\mathbf{H}$, for example Chebyshev polynomials[7]. It helps to think of the powers of **H** on $\mathbf{u}_0$ as generating a sequence of linearly independent states which span (Krylov) subspaces, subsystems through which the system passes as it evolves away from $\mathbf{u}_0$.

This idea of transforming to a one-dimensional evolution can be exploited by the Gram-Schmidt construction of an orthonormal basis, $\mathbf{u}_0, \mathbf{u}_1, \mathbf{u}_2,$ ..., for this sequence of subspaces. Also known as the Lanczos or recursion methods, the construction begins with $\mathbf{u}_0$ and proceeds to construct $\mathbf{u}_{n+1}$ from $\mathbf{H}\mathbf{u}_n$ by removing its component of $\mathbf{u}_n$, $\langle \mathbf{u}_n, \mathbf{H}\mathbf{u}_n \rangle = a_n$, and its component of $\mathbf{u}_{n-1}$, $\langle \mathbf{u}_n, \mathbf{H}\mathbf{u}_{n-1}\rangle = \langle \mathbf{u}_{n-1}, \mathbf{H}\mathbf{u}_n\rangle = b_n$. Since **H** is Hermitian, $\mathbf{H}\mathbf{u}_n$ contains no components of $\mathbf{u}_m$ for m<n-1 (because by construction $\mathbf{H}\mathbf{u}_m$ contains no component of $\mathbf{u}_{n+1}$). So $\mathbf{H}\mathbf{u}_n - a_n\mathbf{u}_n - b_n\mathbf{u}_{n-1}$ is orthogonal to the preceding basis elements, and its matrix element with $\mathbf{u}_{n+1}$ is its normalization, which is also $b_{n+1}$. In the new basis $\{\mathbf{u}_n\}$, **H** becomes a symmetric tridiagonal matrix **T** whose diagonal elements are the $\{a_n\}$ and whose only non-zero, off-diagonal elements are the $\{b_n\}$ on the first two subdiagonals. Since the $\{\mathbf{u}_n\}$ are constructed from powers of **H** acting on $\mathbf{u}_0$, $\mathbf{u}_n$ is a polynomial of degree n in **H**, making $a_n$ and $b_n$ special cases of modified moments, of degree 2n+1 and 2n respectively, optimized by orthonormalization for that particular **H** and $\mathbf{u}_0$.

For numerical applications, $\mathbf{u}_n$ must be localized in the sense that it contains significant contributions from only a finite number of inequivalent orbitals. This in turn depends on $\mathbf{u}_0$ being localized, and on $\mathbf{H}$ being short-ranged or sparse in the sense that each row or column of $\mathbf{H}$ has only a finite number of significant elements. Given that $\mathbf{u}_0$ is localized and $\mathbf{H}$ short-ranged, but not necessarily finite, it follows that $\mathbf{u}_n$ is also localized, although the number of significant components can grow as fast as n-factorial[8].

## 3. Approximations

Starting with 2N+1 optimized moments $\{a_0, a_1, a_2, ..., a_N\}$ and $\{b_0, b_1, b_2, ..., b_N\}$, the central problem is to approximate the imaginary part of $R_0(E)$ for real energies when $\mathbf{H}$ is an infinite matrix. In the classical solution of the moment problem [1], this matrix-element is expressed as the continued fraction,

$$R_0(E) = b_0^2 / E - a_0 - b_1^2 / E - a_1 - b_2^2 / E - a_2 - ... - b_N^2 / E - a_N - b_{N+1} t_{N+1}(E), \quad (4)$$

where $b_0^2$ is the normalization of $\mathbf{u}_0$ and $n_0(E)$ as well as the zeroth power moment, the slash means division by everything to the right, and $t_{N+1}(E)$ is the tail of the continued fraction depending on the rest of the moments, which are not known. This matrix-element $R_0(E)$ has the analytic property that its value and E always lie in opposite halves of the complex plane (upper or lower). This also applies to $t_{N+1}(E)$ and E, and so for E in the lower (upper) half plane, $R_0(E)$ and $t_{N+1}(E)$ are in the upper (lower) half plane. Furthermore, when E has a non-zero, negative (positive) imaginary part, the continued fraction maps the whole half plane of possible values of $t_{N+1}(E)$ into values of $R_0(E)$ within a circle of finite radius $\rho$ which lies entirely in the upper (lower) half of the complex plane. This bound on $R_0(E)$ also bounds $n_0(E)$, and the continued fraction expansion converges in the sense that $\rho$ decreases exponentially with increasing N.

The problem with the continued fraction expansion is that the radius of the bounding circle $\rho$ goes to infinity as E becomes real, which is where $R_0(E)$ must be evaluated for $n_0(E)$. Worse, there is a value of $t_{N+1}(E)$, allowed by the above analyticity requirements, for which the continued fraction in Eq. 4, and hence $n_0(E)$, is infinite; so it is not just the bound on the continued fraction, but the continued fraction itself which goes to infinity on the real E-axis. This is the end of the story unless some additional property of macroscopic systems is used to constrain the values of $n_0(E)$ on the real axis.

## 4. Maximally Broken Time-Reversal Symmetry

Physically, infinities in $n_0(E)$ occur at the energies of states which do not decay, typical of isolated finite systems but not macroscopic systems, for which states generally do decay. This observation offers an alternative approximation to truncation of the continued fraction, an approximation which emphasizes the macroscopic nature of the system by minimizing lifetimes rather than making them infinite. Although it is possible to calculate lifetimes using wave-packets, a simpler approach is to calculate the probability current carried by individual stationary states when the boundary condition at $\mathbf{u}_0$ is altered to make it a source. It is shown in this Sec. that combining a constraint of minimal lifetimes with the moments $\{a_0, a_1, a_2, ..., a_N\}$ and $\{b_0, b_1, b_2, ..., b_N\}$ produces an approximation for the PDoS which converges at each energy at least as fast as the reciprocal of the number of moments, and that the discrete spectrum of a finite system is recovered in the limit where one of the $\{b_n\}$ becomes smaller than the spacing between levels.

In order for a state to decay, its probability must flow away to infinity, and the time-reversed state must have probability flowing in from infinity making this pair of states a time-reversal doublet. In these terms, the condition that states have minimal lifetimes becomes the condition that states break time-reversal symmetry maximally in the sense that probability is carried away at a maximal rate. In precise terms, the state with the maximal rate of probability flow is the state with the largest ratio of probability current to total probability; that is the largest product of probability current, $J$, and normalization $\rho$.

In order to find the states with MBTS, the rate at which probability flows away from $\mathbf{u}_0$ must be calculated by solving Eq. 1. This can be done in many ways, but for numerical work it is convenient to use the tridiagonal basis, $\{\mathbf{u}_n\}$ constructed above. For time-independent $\mathbf{H}$, the time-dependence and spacial dependence of solutions to Eq. 1 can be separated by writing them as $\psi e^{iEt/\hbar}$ where E is the energy of the state and $\psi$ is time-independent so it can be expanded in the $\{\mathbf{u}_n\}$ as $\psi=\psi_0\mathbf{u}_0+\psi_1\mathbf{u}_1+...+\psi_n\mathbf{u}_n+ ...$ , where $\boldsymbol{\psi}$ is the vector whose components are $\psi_0, \psi_1, \psi_2, ....$ In terms of the moments $\{a_n\}$ and $\{b_n\}$ of $\mathbf{H}$, the time-independent Schrödinger equation for this state in the tridiagonal basis $\{\mathbf{u}_n\}$ is the the three-term-recurrence,

$$a_n \psi_n + b_n \psi_{n-1} + b_{n+1} \psi_{n+1} = E \psi_n. \qquad (5)$$

Since the recurrence relates three components of a solution $\psi$, there are two linearly independent solutions for each E, and so a single solution is determined by two conditions on $\psi$, for example specifying $\psi_0$ and $\psi_1$, or some other pair of conditions.

For the stationary state $\psi$ with real energy E, the probability that the system is in $\mathbf{u}_n$ is $\psi_n\psi_n^*$, and the time-derivative of this probability is zero. Since the total probability of a state is conserved for real E, the probability currents into and out of $\mathbf{u}_n$ must be equal. The Schrödinger equation, Eq. 1, relates the time-derivative of a state and the Hamiltonian to give expressions for the probability current $J_{n+1}$ flowing from $\mathbf{u}_n$ to $\mathbf{u}_{n+1}$ which is equal to the current $J_n$ flowing from $\mathbf{u}_{n-1}$ to $\mathbf{u}_n$,

$$J_{n+1} = b_{n+1}\,[\psi_{n+1}^*\,\psi_n - \psi_{n+1}\,\psi_n^*]\,/(i\hbar) = b_n\,[\psi_n^*\,\psi_{n-1} - \psi_n\,\psi_{n-1}^*]\,/(i\hbar) = J_n. \quad (6)$$

Since the current carried between successive $\{\mathbf{u}_n\}$ is independent of n for states with real energies, this is an invariant $J$ of the recurrence in Eq. 5 (related to the Christoffel-Darboux identity [7]). To get a rate of probability flow, this current must be normalized by

$$\rho_N = 1/ (|\psi_0|^2 + |\psi_1|^2 + |\psi_2|^2 + ... + |\psi_N|^2), \qquad (7)$$

(a generalization of the Christoffel function [9]), which is the reciprocal of the total probability that the system is in one of the basis-states $\mathbf{u}_0, \mathbf{u}_1, \mathbf{u}_2, ..., \mathbf{u}_N$. The condition for $\mathbf{u}_0$ to have an extremal lifetime is that the ratio of the probability current $J$ flowing our of $\mathbf{u}_0$, to the total probability in the state,

$$j_N = \rho_N J = [b_1/(i\hbar)] [\psi_1^* \psi_0 - \psi_1 \psi_0^*] / [|\psi_0|^2 + |\psi_1|^2 + |\psi_2|^2 + ... + |\psi_N|^2] \qquad (8)$$

be extremal. The states with extremal rates of probability flow occur in pairs with the two states each carrying the same maximal current, but in opposite directions making them a time-reversal doublet.

The conventional solutions of Eq. 5 are orthogonal polynomials of the first and second kind, $\{p_n\}$ and $\{q_n\}$ respectively, which satisfy the following conditions: $p_0=1/b_0$, $p_1=(E-a_0)/(b_0 b_1)$, $q_0=0$, and $q_1=b_0/b_1$. In terms of these polynomials it is convenient to specify the states $\psi$ with a fixed component, -1, of $\{q_n\}$, and a variable component, z, of $\{p_n\}$, so that,

$$\psi_n = z\, p_n - q_n. \tag{9}$$

Substituting this expression in Eq. 8 gives the rate of probability flow in terms of z,

$$j = [(z^* - z)/(i\hbar)] / [z\, z^*\, P_N - z\, S_N^* - z^*\, S_N + Q_N], \tag{10}$$

where,
$$P_N = |p_0|^2 + |p_1|^2 + \ldots + |p_N|^2, \tag{11}$$

$$S_N = p_0^*\, q_0 + p_1^*\, q_1 + \ldots + p_N^*\, q_N, \tag{12}$$

and,
$$Q_N = |q_0|^2 + |q_1|^2 + \ldots + |q_N|^2. \tag{13}$$

The z which produce extremal values of j occur when the derivatives of j with respect to z and z* are zero,

$$z_\pm = (\mathrm{Re}\{S_N\} \pm i\, [P_N\, Q_N - (\mathrm{Re}\{S_N\})^2]^{1/2})\, /P_N, \tag{14}$$

with the states $\psi^\pm$ generated by $z_\pm$ being the states with extremal positive (+)

and negative (-) rates of probability flow in Eq. 10.

The choice of states $\psi^\pm$ is equivalent to the choice of a tail $t_{N+1}(E)$ for the continued fraction in Eq. 4, as can be seen by dividing both sides of Eq. 5 by $\psi_n$, solving for $b_n \psi_n / \psi_{n-1}$ to get,

$$b_n \psi_n / \psi_{n-1} = b_n^2 / (E - a_n - b_{n+1} \psi_{n+1} / \psi_n), \qquad (15)$$

and linking these expressions to relate $z_\pm$ to $\psi_{N+1}/\psi_N$, which is the tail $t_{N+1}(E)$. This can be shown to have the analytic properties of the tail introduced in the previous Sec., namely that E and $t_{N+1}(E)$ lie in opposite halves (upper and lower) of the complex plane, and that as E goes to infinity, $t_{N+1}(E)$ goes to zero as $1/E$. Now, from Eqs. 15 and 4,

$$R_0(E) = b_0^2 / (E - a_0 - b_1 \psi_1 / \psi_0) = z_\pm, \qquad (16)$$

so the PDoS at E is,

$$n_0(E) = (P_N Q_N - S_N^2)^{1/2} / (\pi P_N), \qquad (17)$$

where $P_N$, $Q_N$, and $S_N$ depend on E and are real, for E real, so that the PDoS is non-negative by the Cauchy-Schwarz's inequality. It preserves the moments $\{a_0, a_1, a_2, ..., a_N\}$ and $\{b_0, b_1, b_2, ..., b_N\}$ because MBTS is equivalent to a choice of a tail t(E) for the continued fraction.

For finite systems, **H** is also finite, the states are localized and do not carry currents to infinity. As a result, the PDoS consists of isolated delta-distributions. A finite **H** transforms to a finite tridiagonal matrix **T** which may be viewed in the context of MBTS as the limit in which one of the off-diagonal matrix-elements, say $b_N$, goes to zero. It is shown in what follows that for $b_N$ much smaller than the separation of zeros of $p_{N-1}$, the approximation for the PDoS in Eq. 17 goes to a sum of isolated delta-distributions located at the zeros of $p_N$, with weights $q_N/p'_N$, where $p'_N$ is the derivative of $p_N$ with respect to E and the expression for the weight is evaluated at the zero. This expression for the PDoS is exactly the same as is obtained by taking $t_N(E)=0$ in Eq. 4, and expressing the finite continued fraction in terms of $p_N$ and $q_N$.

Because $p_N$ and $q_N$ satisfy Eq. 5 with appropriate initial conditions, they each contain a factor of $1/b_N$, so that for $b_N$ very small, they dominate $P_N$, $Q_N$, and $S_N$, in Eqs. 11-13, except close to zeros of either $p_N$ or $q_N$. It is clear from this that the PDoS in Eq. 17 is only non-zero within about $b_N$ of a zero $E_z$ of $p_N$, where $p_N$ can be approximated as $(E-E_z)p'_N$, because $b_N$ is much smaller than the separation of zeros. In Eq. 17, taking $E-E_z$ to be of order $1/b_N$, and discarding those terms in $P_N Q_N - S_N^2$, which are of order $1/b_N$ or smaller, the expression for $n_0(E)$ becomes $(1/\pi)[P_{N-1}(q_N)^2]^{1/2}/[P_{N-1}+((E-E_z)p'_N)^2]$, where all polynomials are evaluated at $E_z$. This is just the expression for a Lorentzian of width $[P_{N-1}]^{1/2}/p'_N$ whose integrated weight is $q_N/p'_N$, the correct value.

## 5. Convergence

For there to be a time-reversal doublet and hence a decaying state at energy E, there must be some linear combination of $\mathbf{p}_N$ and $\mathbf{q}_N$ which carries a current to infinity as N goes to infinity. This requires that both $\mathbf{p}_N$ and $\mathbf{q}_N$ be spread evenly over the $\{\mathbf{u}_n\}$, and that the angle $\theta_N$ between them not go to zero; resulting in $P_N$, $Q_N$, and $S_N$ which all increase linearly with N at the same rate. The consequence of this is that when there is a time-reversal doublet at energy E, the expression for $n_0(E)$ in Eq. 17 converges point-wise as $1/N$.

The other case where Eq. 17 converges point-wise is when E is in a gap in the PDoS. Since there is no physical state at this energy, the $\{p_n\}$ diverge with increasing n, and so $P_N$ also diverges with increasing N. $Q_N$ cannot diverge any faster than $P_N$ because the current in Eq. 6 must be conserved when $\psi_n$ is $p_n+iq_n$, but angle $\theta_N$ goes to zero because at least the $\{p_n\}$ diverge. The result of this is that for an energy in a gap, the PDoS in Eq. 17 converges to zero at the same rate that $P_N$ diverges, which is exponentially in most cases.

The remaining cases are where E is either the energy of a state which belongs to a time-reversal singlet, or an accumulation point of such states (an infinite number of such states differ by arbitrarily small energies). At these energies the PDoS is singular in the sense that its value can be changed between zero and infinity with arbitrarily small changes in the Hamiltonian. Since the approximation in Eq. 17 produces a smooth PDoS for N finite, $n_0(E)$ approaches the singular spectrum by becoming rougher with increasing N. The PDoS converges in the mean (integrals over smooth functions converge) at these energies, rather than point-wise.

The behavior of this approximation based on MBTS is illustrated in Fig. 1 for a spectrum consisting of two semi-elliptical bands with a gap between them. The approximation in Eq. 17 using 10, 22, and 46 moments are shown as the full lines a, b, and c, with the full line d being the exact PDoS. The values of $n_0(E)$ are displaced by 0.2 for b, 0.4 for c, and 0.6 for d. As required, the approximate densities cross the exact density a number of times equal to the number of moments used. This result is somewhat better than the results of previous methods for terminating continued fractions [10].

Maximizing the entropy functional, $\int n_0(E) \text{Ln}[n_0(E)]\, dE$, is a general approach to approximation[11], and is related to this work in the sense that states with MBTS decay fastest, and therefore generate entropy fastest. For 2N moments, the PDoS with maximal entropy is the exponential of a polynomial of degree 2N whose coefficients are chosen to reproduce the normalization and given moments[12]. While the current density carried by a state is expressed above as a ratio of quadratics, and so can be maximized analytically, maximizing the entropy functional can only be done numerically and the implementation of this used here suffers from instabilities.

For comparison, the dotted curves a and b are the maximum entropy [13] approximations for 10 and 22 moments respectively; the program used did not converge for 46 moments. There is no doubt that a solution with maximum entropy exists, the problem is lack of an algorithm which always solves this strongly non-quadratic optimization. The differences between maximal entropy and MBTS are clear even from this limited comparison. The main one is that the maximum entropy approximation to the PDoS decreases exponentially in E outside the bands, while MBTS approximation decreases algebraically, as some inverse power of E outside the bands. Since the PDoS is normalized to one, the smaller weight of the maximum entropy approximation outside the bands, produces a greater weight inside the bands resulting in a smaller error. The amplitudes of the oscillations inside the bands are similar.

The point-wise convergence of the PDoS is demonstrated in Fig. 2 for the two-band model shown in Fig. 1. Lines a, b, and c in Fig. 2a show for increasing numbers of moments the convergence of the value of the PDoS at energies: -2.5, -0.5, and 2.5, which are outside the bands and in the gap. The straight lines which result from plotting the log error against level number (half the number of moments used) show that the convergence is indeed exponential as argued above. These calculations were done in double precision arithmetic, so the exponential decrease in the error stops when it reaches the rounding error. In Fig. 2b, curves a, b (the very jagged curve), and c show for increasing numbers of moments the convergence of the value of the PDoS at energies: -2.0, -1.5, and 0.0, which are at band-edges and inside a band. For the band-edges, curves a and c, the error decreases a little faster than the reciprocal of the number of levels. It is a little surprising that the error at these singular points decreases faster than the error for an energy inside one band, curve b, where despite large downward fluctuations in the error, the maximum error decreases as the reciprocal of the number of levels. The difficulty in converging entropy maximizations makes it impossible to compare with these results.

## 6. A Single-band Approximation

The simplest time-reversal doublet which satisfies MBTS is the pair of plane-waves, $\psi_n = \exp\{\pm in\phi\}$, where $\phi$ is some real angle between 0 and $\pi$. For a finite band, these are solutions to Eq. 5 for n large when $a_n$ and $b_n$ are constants or go to constants for large n [7]. Similarly, for an infinite band, these are solutions for n large when the ratios $a_n/b_{n+1}$ and $b_n/b_{n+1}$ become constant with increasing n.

Knowing that the MBTS states are asymptotically plane waves makes it much simpler to determine the PDoS, namely that for large N,

$$\psi_N / \psi_{N-1} \approx \psi_{N+1} / \psi_N, \tag{18}$$

at each energy E which is equivalent to the condition that the tails have converged, $t_N(E) = t_{N+1}(E)$. Substituting for the (N-1)-th, N-th, and (N+1)-th components of $\psi$ from Eq. 9 produces a quadratic equation for $z_\pm$ whose solutions are,

$$z_\pm = [B_N \pm \sqrt{(B_N^2 - 4 A_N C_N)}]/(2 A_N), \tag{19}$$

where,

$$A_N = p_N^2 - p_{N+1} p_{N-1}, \tag{20}$$

$$B_N = 2 p_N q_N - p_{N+1} q_{N-1} - p_{N-1} q_{N+1}, \tag{21}$$

and,

$$C_N = q_N^2 - q_{N+1} q_{N-1}. \tag{22}$$

As in the case of the more general application of MBTS, the approximate PDoS has the given moments because the approximation can be expressed as a boundary condition on the tail of the continued fraction.

As an example of this approximation, consider the electronic structure of squarium, a model material consisting of tight-binding s-orbitals on a square lattice with hopping between nearest neighbors only. The exact PDoS has discontinuities (imaginary logarithmic singularities) at the edges of the band and goes to infinity logarithmically (a real logarithmic singularity) at the center of the band. Although the model has only one band of electronic states, the two kinds of logarithmic singularities challenge many approximation methods. Figure 3 shows the above approximation for 10, 22, 46, and 400 moment as full curves a, b, c, and d respectively. The dotted versions of curves a and b are the maximum entropy approximations for 10 and 22 moments respectively; the maximum entropy method did not converge for 46 or 400 moments.

      The results in Fig. 3 show that for 10 and 22 moments, the single-band approximation has smaller oscillations throughout the band and sharper band-edge discontinuities than maximum entropy, which in turn has smaller oscillations and sharper edges than the more general MBTS from Sec. 4, not shown. Maximum entropy does slightly better than the single-band approximation on the singularity at the center of the band, and not quite as well at the band edges. The sharpness of the band-edges in the single-band approximation is surprising because, like the more general MBTS, it is algebraic outside the band, but in this case crosses the exponential maximum entropy approximation where both are too small to see on this scale.

      This single-band approximation should also be compared with the constant terminator of earlier work [10] where for finite single bands, the asymptotically constant $\{a_n\}$ and $\{b_n\}$ were replaced by constants $a_n$=a and

$b_n=b$. This also produced a quadratic equation for $t_N(E)$, but one in which the coefficients were either constant or linear in E, not of degree 2N in E as they are in the quadratic resulting from Eq. 18. Requiring the tail to be converged rather than the moments constant produces a remarkable improvement in the approximation as can be seen by comparing the results in Fig. 3 with similar results for constant terminators in Ref. 10.

## 7. Approximate Greenians

The PDoS is proportional to a singular part of the more general complex function of E, the Greenian or resolvent,

$$R_0(E) = <\mathbf{u}_0, (E - H)^{-1} \mathbf{u}_0>. \tag{23}$$

Part of this function, the physical sheet, can be expanded for complex E as the continued fraction in Eq. 4, and can have singularities only at real energies. If the Hamiltonian H permits states which break time-reversal symmetry, time-reversal doublets, $R_0(E)$ can be analytically continued to a second sheet which contains additional singularities corresponding to decaying states, resonances and unstable states (see Ref. 6 for an example). It is frequently useful to know the complex energies of these second-sheet singularities, and in principle, the moments determine these as well as the PDoS.

While Sec. 4 MBTS was used to construct an approximation for the PDoS at real energies, the same argument can be extended to complex energies. The expressions for the current and the normalization, Eqs. 6 and 7, are already in their complex form as is Eq. 14, which in Eq. 16 is shown to be equal to $R_0(E)$. For complex energies, $z_-$ in Eq. 14 gives a good approximation to the physical sheet of the Greenian for E in the upper half of the complex E-plane, and $z_+$ is a good approximation in the lower half-plane. While they are good approximations on the physical sheet, $z_+$ and $z_-$ in Eq. 14, cannot be analytically continued to the second sheet because they are functions of E and E*, not just E.

While the formulation of MBTS in Sec. 4 is not analytic in E, the single-band formulation in Sec. 6 is, because the condition that states go to pure exponentials, Eq. 18, does not involve complex conjugates. The expression for $z_\pm$ in Eq. 19 can be evaluated for complex E and produces both the physical and second sheets of $R_0(E)$. As a numerical example this approximation is applied to the PDoS $\text{sech}(\pi E)$ which arises in classical diffusion [6]. This is a case where the PDoS extends to infinite energy, but it satisfies the criterion for a single band in Sec. 6. While the this PDoS consists of a single broad peak centered at zero energy, the second sheet of the Greenian which produces this PDoS has poles at $\pm i$, $\pm 2i$, …

Along the imaginary E-axis, $z_-$ is purely imaginary because the PDoS is symmetric about E=0; and this is plotted at 301 points for 10, 22 and 46 moments, a b and c, respectively. The first two of the diffusion poles along the negative imaginary E-axis are resolved with all approximations doing well for the pole at -i and the approximations with more moments doing better for the pole at -2i. None of the approximations show any sign of the pole at -3i, and including additional moments in the approximation does not lead to additional poles being resolved. The reason is that at complex energies the magnitudes of the polynomials $\{p_n\}$ and $\{q_n\}$ increase rapidly with n. The condition that $\psi_n$ goes to a pure exponential relates $\{p_n\}$ and $\{q_n\}$ for large n, while the value of $z_\pm$ is just $\psi_0$, so the value of the Greenian depends on cancellations between large quantities to get a small one. At the value of N for which $1/p_N$ or $1/q_N$ becomes less than the rounding error, the approximation becomes insensitive to further moments, and the region of the second sheet which can be approximated is therefor limited by the rounding error.

## 8. Quadratic Approximations

It is simplest to consider the Greenian or resolvent in terms of the three-term recurrence in Eq. 5. For each value of E, a solution $\{\psi_n\}$ is constructed in Sec. 4 from a single complex number z, with the normalization and phase of the solution fixed by the relation between $\psi_0$ and $\psi_1$. For energies in the upper half of the complex energy-plane, it is clear that there are two special solutions to the recurrence: one $\psi^-$ for which $\psi_n^-$ goes to exponentially to zero as n goes to infinity; and another $\psi^+$ for which $|\psi_n^+|$ increases faster with n than any other solution.

Of these two special solutions, the most convergent $\psi_n^-$ and the most divergent $\psi_n^+$, it is the latter which is difficult to determine numerically because its dependence on z becomes weaker and weaker with increasing n, in contrast with the former where the dependence on z increases with n. Defining $z_+$ and $z_-$ more generally than before, as the initial conditions which generate $\psi^+$ and $\psi^-$ respectively, $z_-$ is the value of the Greenian on the physical sheet and $z_+$ is its value on the second sheet, in the upper half of the energy-plane. On the real energy-axis, they become the two solutions defined in Sec. 4, and in the lower half-plane they interchange sheets becoming respectively the second and physical sheets.

If both the physical and second sheets of the Greenian are to be approximated, then it is reasonable to expect these approximations to be solutions to a quadratic equation, and indeed that is consistent with the above applications of MBTS. In Sec. 4 the quadratic equation resulted from minimizing the ratio of current to total probability giving a result which applies only on the physical sheet because these depend on both $\psi$ and $\psi^*$. In Sec. 7, the quadratic equation arises in the single-band approximation from the condition that the solution to the recurrence becomes a pure exponential, which can be analytically continued because it depends only on $\psi$ and not on $\psi^*$.

When there are gaps in the spectrum, as for example in Fig. 1, the asymptotic ratios of moments, $a_n/b_{n+1}$ and $b_n/b_{n+1}$ are generally quasi-periodic in n[14, 15] in the same sense that these ratios are asymptotically constant for a single band. In exceptional cases such as bands of equal width they are asymptotically periodic in n. In these latter cases where for large n, there is some M such that, $a_n/b_{n+1} \approx a_{n+M}/b_{n+M+1}$ and $b_n/b_{n+1} \approx b_{n+M}/b_{n+M+1}$, the states with MBTS are not asymptotically pure exponentials, but rather Bloch states,

$$\psi_n^+ \approx \exp\{i n \phi\} u(n), \tag{24}$$

for which $u(n) \approx u(n+M)$ for large n, with $\phi$ some angle between 0 and $\pi$; $\psi_n^-$ is $(\psi_n^+)^*$. Now the condition for the MBTS solutions is,

$$\psi_N / \psi_{N+M} \approx \psi_{N+m} / \psi_{N+M+m}, \tag{25}$$

where $0 < m < M$. In the case when m=1, this is equivalent to requiring that the tails have converged, $t_N(E) = t_{N+M}(E)$. As in the single-band case, this approximation for asymptotically periodic ratios of moments gives an excellent PDoS and Greenian. The general case where the moments are quasi-periodic remains to be investigated.

Properties of electronic systems, such a conductance, bond order, and so forth, depend on correlations between different components of states and so are related to off-diagonal elements of $(E-H)^{-1}$ rather than its diagonal elements which give the PDoS as in Eq. 23. In these cases, the various moments become matrices whose rows and columns correspond to the correlated components; the continued fraction in Eq. 4 becomes a matrix continued fraction; and the recurrence in Eq. 5 becomes a matrix recurrence whose solutions are linear combinations of matrix orthogonal polynomials [16, 17]. It seems likely that the idea of MBTS can be extended to these problems.

## 9. Summary and Conclusions

In contrast to the states of finite systems which are spacially localized and so contribute discrete delta-distributions to the density of states, the states of macroscopic systems are typically extended and current carrying so that they contribute smooth bands to the density of states. This observation is used to develop approximations for the densities of states of macroscopic systems using states which carry current to infinity at a maximal rate, or in other words, states which maximally break time-reversal symmetry. The advantage of these approximations is that the rate at which states carry current to infinity can be maximized independently at each energy, and this rate is a ratio of quadratics whose maxima can be found analytically. Maximizing the current is related to maximizing the entropy functional of the density of states, but this latter method has the disadvantage that it is global and non-quadratic, so that its solution cannot always be found with a given algorithm.

The principle of MBTS is applied to approximating densities of states given their moments, modified moments, or tridiagonal representation. When nothing else is known about the structure of the states, this leads to an approximation which converges inversely with the number of moments, although it is inferior to maximizing entropy in the cases where this is possible. When the states are known to be asymptotically plane-waves or Bloch states, MBTS leads to an approximation which is comparable to maximizing entropy, but always converges. As well as the density of states, this approximation converges rapidly for the first sheet of the Green function, and converges asymptotically for the second sheet of the Green function.

**Acknowledgments**

The Authors are grateful to the Richmond F. Snyder Fund and the University of Oregon Foundation for support of this work, to David Drabold for advice about maximum entropy methods, to Larry Bretthorst for the use of his program, and to Geoff Wexler for reading drafts of this paper. RH thanks the Theory of Condensed Matter group of the Cavendish Laboratory and Pembroke College Cambridge for their hospitality during some of this work.

**Figure Captions**

1. A comparison of MBTS (full line) and maximum entropy[13] (dotted line) approximations to a two-band model PDoS using 10, 22, and 46 moments for curves a, b, and c, respectively. Successive curves are displaced vertically by 0.2, and curve d (full line) is exact.

2a. Demonstration of point-wise convergence of the MBTS approximation for the example in Fig. 1 at energies outside the bands: -2.5 (a), -0.5 (b), and 2.5 (c). The number of moments is twice the number of levels.

2.b. Demonstration of point-wise convergence of the MBTS approximation for the example in Fig. 1 at energies on the band edges, or inside the band: -2.0 (a), -1.5 (b), and 0.0 (c). The number of moments is twice the number of levels.

3. A comparison of the single-band MBTS (full line) and maximum entropy[13] (dotted line) approximations for the PDoS of squarium, using 10 and 22 moments, curves a and b respectively, and just the single-band MBTS for 46 and 400 moments, curves c and d, respectively. Successive curves are displaced vertically by 0.2.

4. Single-band MBTS approximations for the physical and second sheets of the Greenian corresponding to the PDoS sech($\pi E$), plotted along the imaginary E-axis at 301 points using 10 (a), 22 (b), and 46 (c) moments.

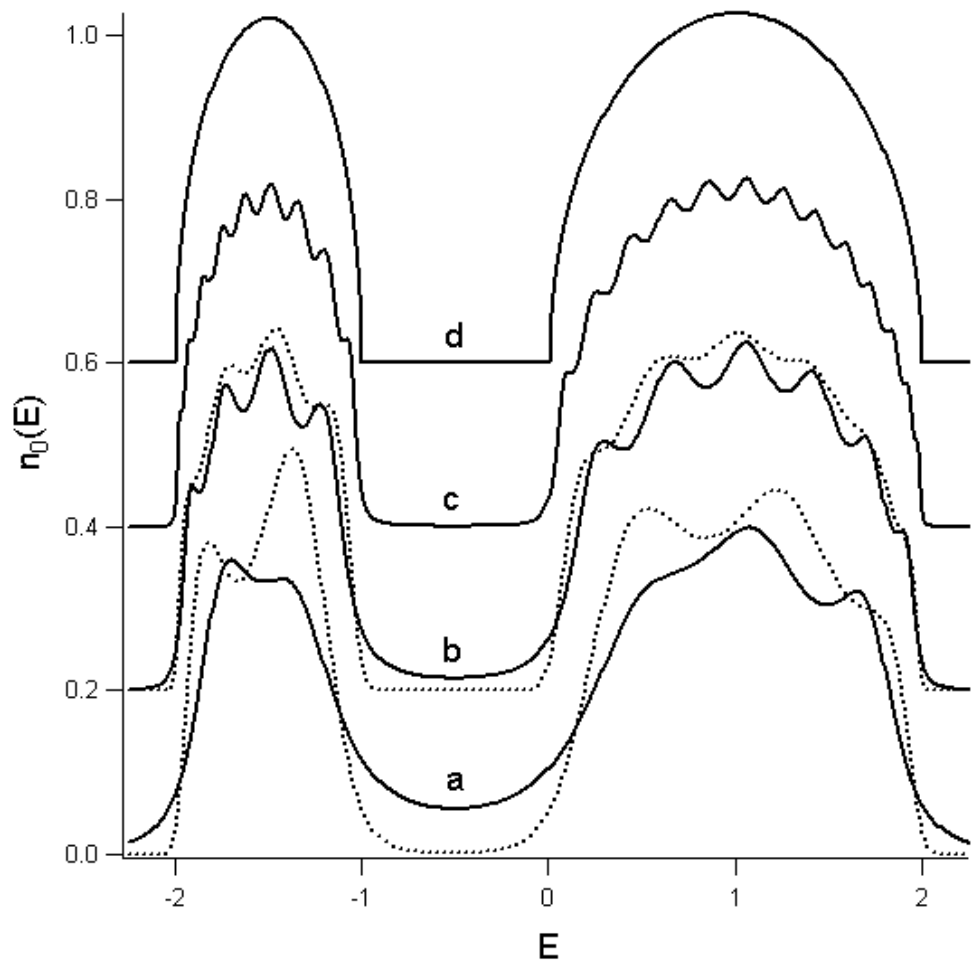

Figure 1.

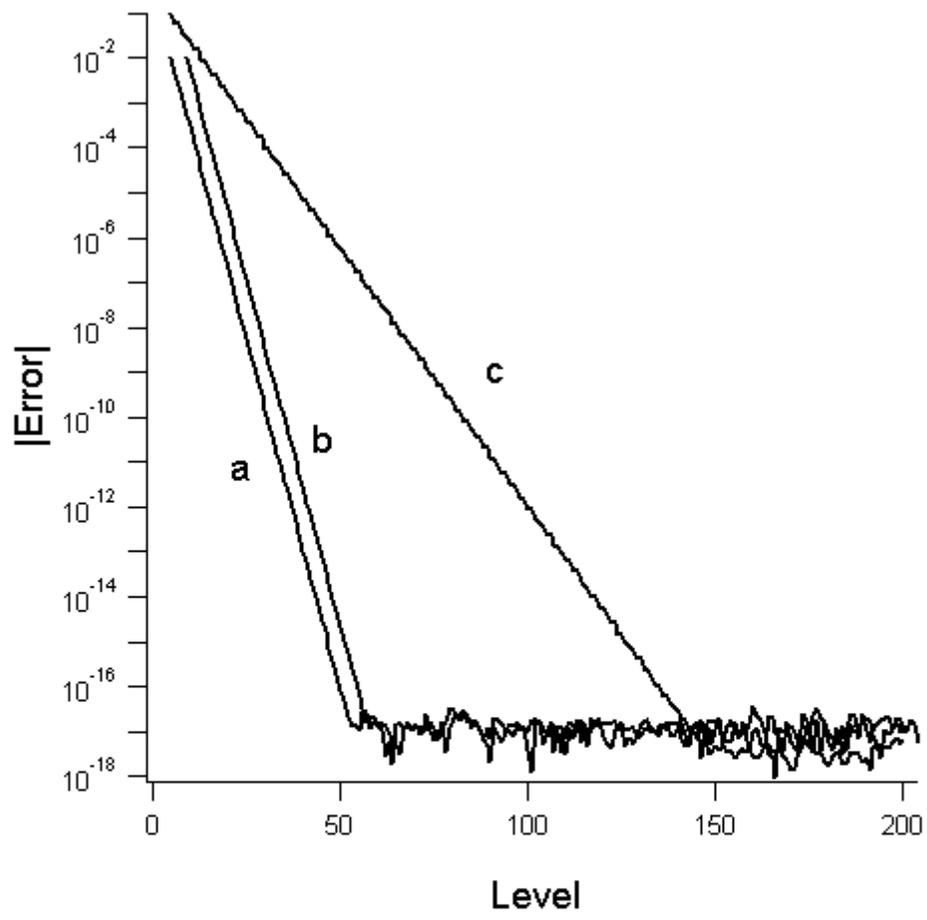

Figure 2a.

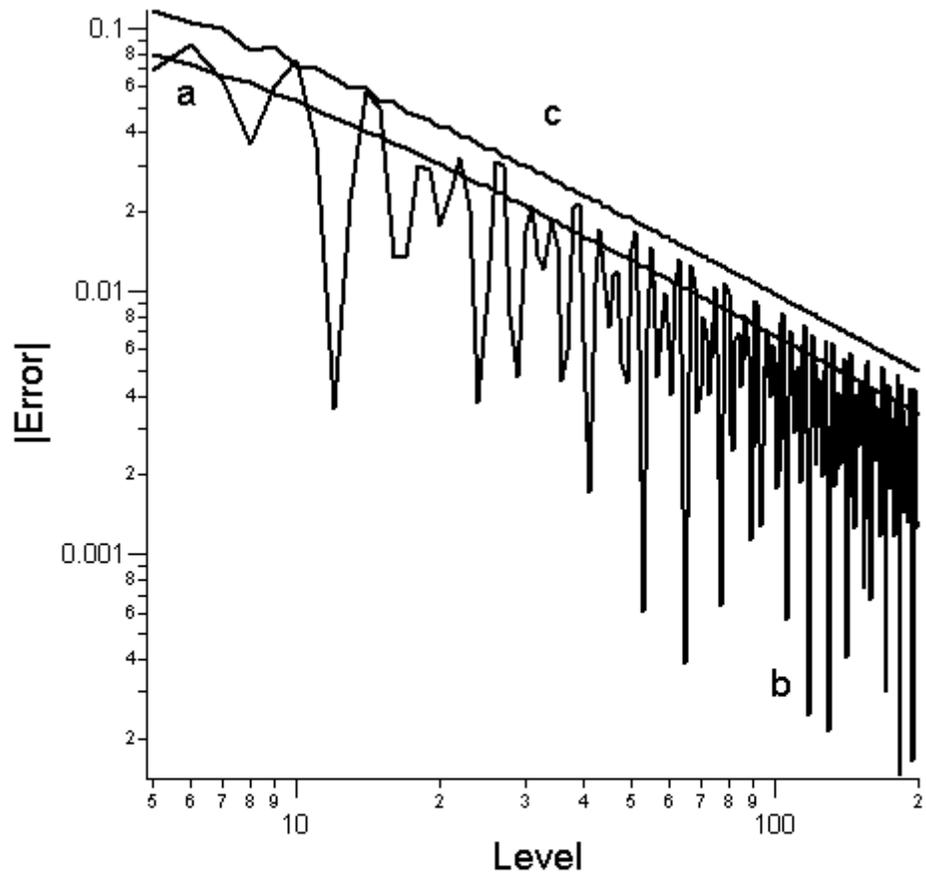

Figure 2b.

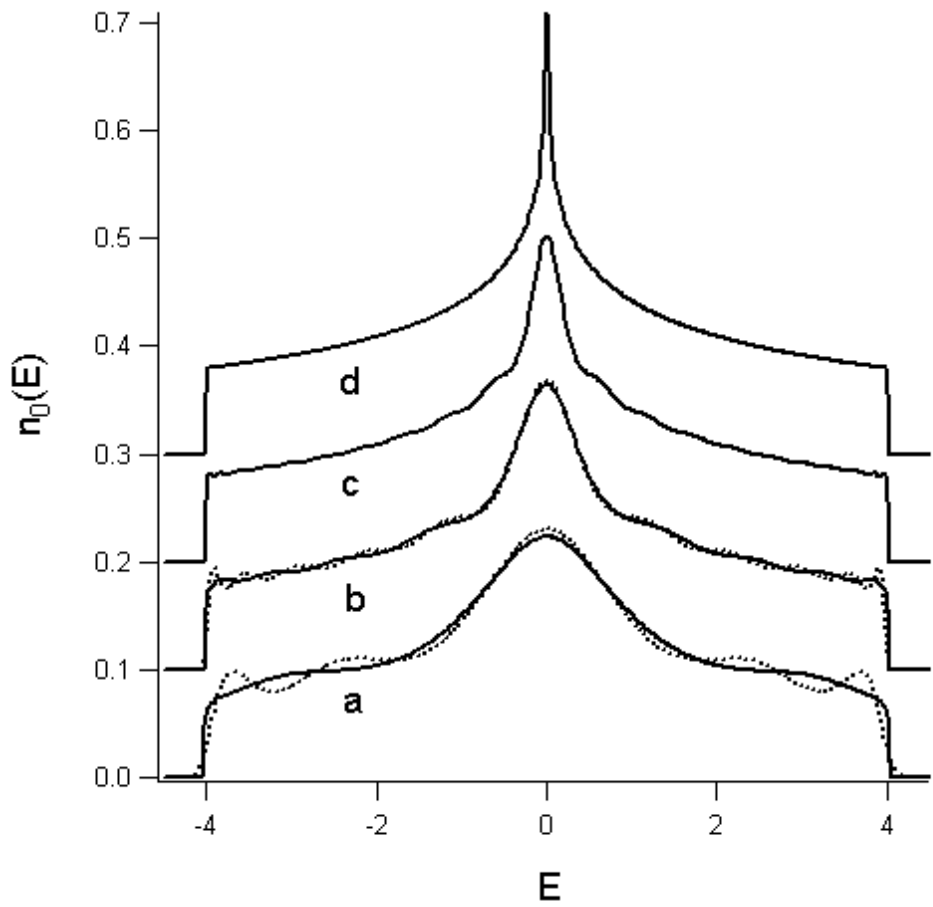

Figure 3.

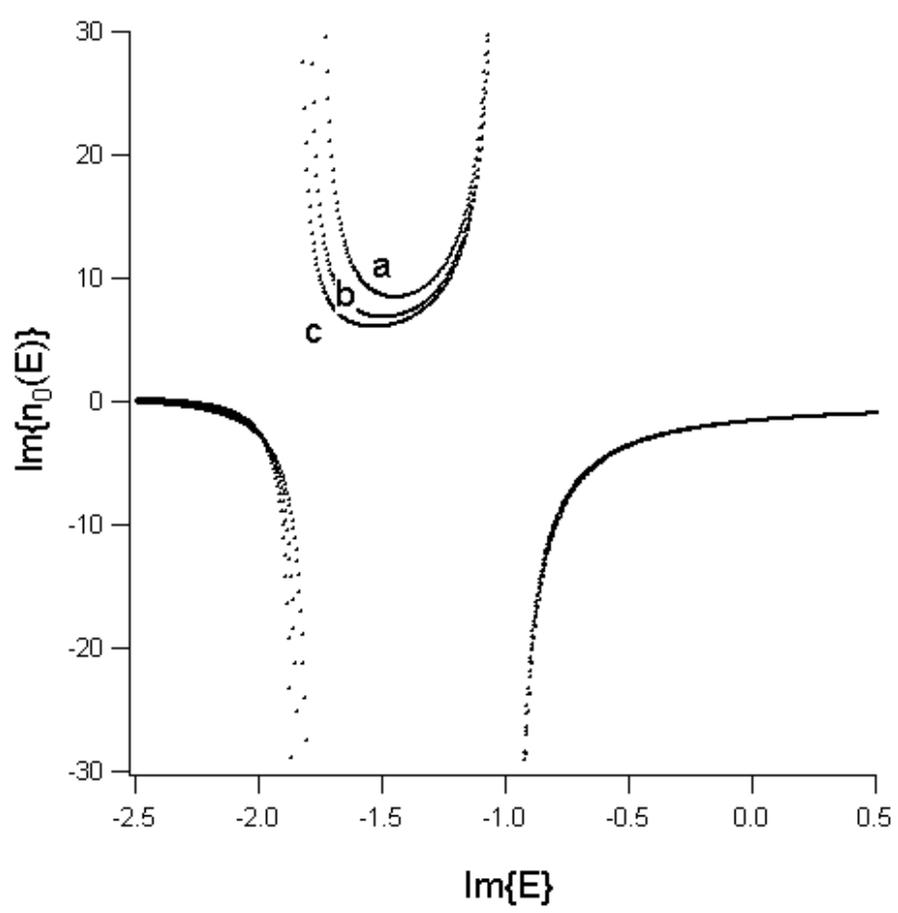

Figure 4.